\documentclass[12pt]{article}
\usepackage{amsmath}
\usepackage{amssymb}
\usepackage{amsthm}
\usepackage{amsfonts}
\usepackage{latexsym}
\usepackage{varioref}
\usepackage{ifthen}
\usepackage{epsf}
\usepackage{epsfig}
\usepackage{graphicx}
\usepackage{color}
\usepackage{subfig}
\usepackage{booktabs}
\usepackage{multicol}
\usepackage{float}
\begin{document}
\parindent 0mm 
\setlength{\parskip}{0.5\baselineskip} 
\thispagestyle{empty}
\pagenumbering{arabic} 
\setcounter{page}{0}

\mbox{ }
\hfill UCT-TP-282/10
\newline
\mbox{ }
\hfill MZ-TH/10-40
\newline
\mbox{ }
\hfill \today
\newline
\begin{center}
{\Large \bf 
Confronting electron-positron annihilation into hadrons  
with QCD: an operator product expansion analysis\footnote{
\footnotesize 
Supported in part by  the National Research Foundation (South Africa), and the Alexander von Humboldt Foundation (Germany).}
}
\end{center}
\begin{center}
{\bf S. Bodenstein}$^{(a)}$, 
{\bf C. A. Dominguez}$^{(a),(b)}$, 
{\bf S. I.  Eidelman}$^{(c),(d)}$,
\\
{\bf H. Spiesberger}$^{(e)}$,
{\bf K. Schilcher}$^{(e)}$
\end{center}
\begin{center}
$^{(a)}$Centre for Theoretical \& Mathematical Physics,
\\ 
University of Cape Town, Rondebosch 7700, South Africa
\\
$^{(b)}$Department of Physics, Stellenbosch University, 
Stellenbosch 7600, South Africa
\\
$^{(c)}$ Budker Institute of Nuclear Physics, 630090, 
Novosibirsk, Russian Academy of Science, Russia
\\
$^{(d)}$ Novosibirsk State University, 630090, Novosibirsk, Russia
\\
$^{(e)}$ Institut f\"{u}r Physik, Johannes Gutenberg-Universit\"{a}t
Staudingerweg 7, D-55099 Mainz, Germany
\end{center}

\begin{center}
\textbf{Abstract}
\end{center}
\noindent
{\small 
Experimental data on the total cross section of $e^+ e^-$ 
annihilation into hadrons are confronted with QCD and the 
operator product expansion using finite energy sum rules. 
Specifically, the power corrections in the operator product 
expansion, i.e. the vacuum condensates, of dimension $d = 2$, 
$4$ and $6$ are determined using recent isospin $I=0+1$ 
data sets. Reasonably stable results are obtained which are 
compatible within  errors with values from $\tau$-decay. 
However, the rather large data uncertainties, together with the 
current value of the strong coupling constant, lead to very large 
errors in the condensates. It also appears that the separation into 
isovector and isoscalar pieces introduces additional uncertainties 
and errors. In contrast, the high precision $\tau$-decay data of 
the ALEPH collaboration in the vector channel allows for a more 
precise determination of the condensates. This is in spite of QCD 
asymptotics  not quite been reached at the end of the $\tau$ 
spectrum. We point out that isospin violation is negligible in 
the integrated cross sections, unlike the case of individual 
channels.
}


\newpage
\bigskip
\noindent


\section{Introduction}
In the framework of QCD sum rules \cite{SVZ}-\cite{REV}, the quark and gluon 
propagator modifications due to confinement are parametrized in 
terms of vacuum expectation values of quark and gluon fields. 
These so-called vacuum condensates appear as power corrections 
to perturbative QCD in the operator product expansion (OPE) 
of current correlators at short distances. The QCD condensates 
are accessible from experiment through the spectral functions, 
i.e.\ the imaginary parts of the current correlators. Since from 
dimensional arguments there are no gauge invariant operators of 
dimension $d=2$ built from quark and/or gluon fields in QCD, it 
has been customary to assume these power corrections to start at 
$d=4$. However, one cannot exclude a priori some dynamical 
mechanism capable of generating a term of $d=2$ in the OPE 
\cite{Chetyrkin:1998yr}. In a recent paper \cite{Lee:2010hd} it 
has been argued that there is no evidence for such a term in 
lattice QCD, contrary to earlier claims. There is also no 
definite evidence from $\tau$-decay as shown in  recent  sum 
rule analyses \cite{C2_1}-\cite{C2_4} 
based on the final ALEPH data \cite{ALEPH2}. These determinations 
have also found values for the dimension $d=4$ and $d=6$ 
condensates consistent with expectations. 

Given the impact of these condensates on QCD sum rule 
applications it is important to attempt a determination 
based on experimental information independent of $\tau$-decay. 
Historically, determinations based on $e^+ e^-$ annihilation 
data have preceded $\tau$-decay analyses \cite{epluseminus}. 
Given the current availability of more accurate data from 
$e^+ e^-$ annihilation, and the sizable increase in the value 
of the strong coupling over the years, it is imperative to 
perform a fully updated determination of the condensates to 
confront with results from $\tau$-decay. An additional motivation 
for such a project is related to the current discrepancy in the 
hadronic vacuum polarization value of the $g-2$ factor of the 
muon obtained from $\tau$-decay and from $e^+ e^-$ annihilation 
data \cite{davier2011}, to wit.
The Standard Model (SM) predictions of $a_{\mu}$, the anomalous 
magnetic moment of the muon, are limited in precision by 
contributions from the hadronic vacuum polarization. The dominant 
terms are usually calculated via a weighted dispersion relation 
involving experimental cross section data on $e^{+}e^{-}$ 
annihilation into hadrons at low energies, and perturbative 
QCD (PQCD) at high energies. The required integration kernel 
emphasizes low photon virtualities. During the last decade the 
SM prediction of $a_{\mu}$ has improved continuously as more 
precise experimental data became available \cite{G2}. Nonetheless, 
a discrepancy between the measured and the predicted values remains 
and it is tempting to trace this discrepancy back to {\it new 
physics} beyond the SM. Unfortunately, the experimental hadronic 
data entering the analysis originate from different experimental 
groups and often errors do not overlap. In addition, most of the 
modern experiments are dominated by systematic uncertainties which 
are difficult to estimate. Consequently, one should test the data 
base to the greatest possible extent against the underlying 
theory, i.e.\ QCD, before accepting the drastic conclusion that 
the SM has failed to describe the data. 

In the present paper we attempt to investigate the 
status of experimental $e^+e^-$ annihilation data by confronting 
them with the OPE using QCD sum rules. We study the first three moments $M_N(s_0)$ 
($N=0$, 1, 2) of $R(s)$, the ratio of the cross sections for 
hadron to muon production in $e^+e^-$ annihilation, at a finite 
upper limit $s_0$ of the integration over the square of the 
center-of-mass energy, $s_0$. These moments are related to the 
QCD condensates of dimension $d=2N+2$. The extracted values of 
these condensates are very sensitive to experimental and theoretical 
errors as they result from evaluating a small difference between 
two large numbers, i.e.\ the difference between integrated data 
and PQCD. The sum rules therefore constitute a sensitive test of 
the consistency of the data with general expectations from QCD. 
In parallel with the determination of a potential $d=2$ term in 
the OPE, we also consider power corrections of dimension $d=4$ 
and $d=6$. This can help establish whether there is evidence for 
a dynamical $d=2$ power correction, and whether the results for 
the higher dimensional condensates are consistent with 
expectations. There is also independent information on the $d=4$ 
gluon condensate from charmonium analyses, as well as theoretical 
estimates of the $d=6$ four-quark condensate which can be used to 
check for consistency. We limit the analysis to dimension $d=6$, 
as higher dimensional condensates are affected by such large 
uncertainties that no meaningful value can be obtained at present.

In the evaluation of the moments, the upper limit of the sum rule 
integral must be taken large enough for PQCD to be applicable. It 
should be recalled in this connection that the QCD asymptotic regime 
is reached in the time-like region only at energies much higher 
than in the space-like region. For instance, this is seen in 
$\tau$-decay data of the ALEPH collaboration \cite{ALEPH2}, where 
the $V-A$ hadronic spectral function does not agree with the 
vanishing result of PQCD even at $s_{0}\approx m_{\tau}^{2}$. 
In contrast, deep inelastic electron-nucleon scattering is well 
known to exhibit precocious scaling. 

In connection with the determination of the muon magnetic moment 
anomaly, $a_{\mu}$, only the sum of the isospin $I=1$ and $I=0$ 
spectral functions is needed. However, QCD also makes definite 
predictions for the moments of the separate isovector and 
isoscalar components of the spectral functions. The corresponding 
moments are related by a trivial overall factor determined by the 
quark charges. Therefore, from the theoretical point of view it 
is interesting to study both the isovector and the isoscalar 
channels of $R(s)$ separately. Of specific interest is the $I=1$ 
channel since independent data are available in this case from 
hadronic $\tau$-decay. For example, it was suggested in  
\cite{Alemany:1997tn} to use the $\tau$-spectral functions in the 
calculation of the  muon anomaly. To be precise, the authors of 
\cite{Alemany:1997tn} replaced the $\tau \rightarrow \nu_{\tau} + 
\pi^{+} \pi^{0}$, $2\pi^{-} \pi^{+} \pi^{0}$ and $\pi^{-}3\pi^{0}$ 
spectral functions measured in $\tau$-decay with the poorly known  
neutral $2\pi$ and $4\pi$ spectral functions from $e^{+}e^{-}$ 
annihilation. For these exclusive final states, sizable 
model-dependent isospin-breaking effects need to be incorporated. 
This is in contrast to the case of moments, as opposed to 
individual channels, where isospin breaking is negligible for 
large enough $s_0$. But if the aim is to determine QCD condensates 
for definite isospin, then the necessary separation of $e^+e^-$ 
data into final states with $I=0$ or $I=1$ would require a 
detailed understanding of isospin violation. This separation, 
though, would not even be possible in the presence of mixing. 
This mixing has been known for quite some time to be sizable 
in the $\rho$-$\omega$ system. In addition, it was shown recently 
that $\rho$-$\gamma$ mixing could, perhaps, explain the well-known 
discrepancy between $e^+e^-$- and $\tau$-decay-based evaluations 
of $a_{\mu}$ \cite{Jegerlehner:2011ti}. However, for a determination 
of QCD condensates one does not need such a detailed, and generally 
model dependent analysis. In fact, isospin is almost exactly 
conserved in integrated cross sections or moments, provided the 
upper limit of the integration over squared energies energies, $s_0$, is taken 
large enough. This is because isospin-breaking terms are 
proportional to $m_q^2 / s_0$ where $m_q$ refers to the mass 
of the light quarks. Therefore one can restrict the analysis 
of moments and condensates to the total ($I=1$ plus $I=0$) 
$e^{+}e^{-}$ cross section. 

A similar analysis based on the known $I=1$ spectral functions 
obtained from $\tau$-decay data has already been performed 
\cite{C2_1}-\cite{C2_4}. However, 
there the upper limit of integration $s_0$ is restricted by the 
$\tau$ mass to values of about $s_0 \simeq 3\, \mbox{GeV}^2$ and 
it is not clear whether this value is large enough for PQCD to 
be applicable. In spite of this, the condensates derived from 
$\tau$-decay have been shown previously to be consistent with 
QCD expectations and we will therefore focus in the following 
on the case of $e^{+}e^{-}$ annihilation.

The paper is organized as follows. In Section~2 we present a
short summary of the theoretical background and in Section 3 
we describe shortly the data used in our analysis, with a more 
detailed discussion  deferred to the Appendix. Section 4 contains 
our results, their interpretation, and our conclusions.


\section{QCD Sum Rules}

We begin by considering the electromagnetic current correlator
\begin{eqnarray}
\Pi^{\rm EM}_{\mu\nu}(q^2) & = &\;
i\int d^{4}x \; e^{iq x}\;
\langle 0|T(J^{\rm EM}_\mu(x)\, J^{\rm EM}_\nu(0))|0\rangle 
\nonumber 
\\[.2cm]
& = &
\left(  -q^{2}g_{\mu\nu}+q_{\mu}q_{\nu}\right) \, 
 \Pi^{\rm EM}(q^{2}) \, ,
\end{eqnarray}
where for three flavours 
\begin{equation}
J^{\rm EM}_\mu(x) = 
\frac{2}{3} \bar{u}(x)\gamma^{\mu}u(x) 
- \frac{1}{3} \bar{d}(x)\gamma^{\mu}d(x) 
- \frac{1}{3} \bar{s}(x)\gamma^{\mu}s(x) \, .
\end{equation}
QCD is flavour blind, and if isospin invariance 
is exact, it is convenient to define a QCD current correlator using 
any of the quark currents $\bar{q}_i \gamma_{\mu} q_i$ with flavour 
$i$. This leads to
\begin{equation}
\operatorname{Im} \Pi^{\rm EM}(q^2) = \sum_{i=1}^{n_f} Q_i^2
\operatorname{Im} \Pi_{\rm VV}(q^2)\; , 
\end{equation}
where $Q_i$ is the charge of the quark $i=u$, $d$, $s, \ldots$, 
$\sum_i Q_i^2 = 2/3$ for $n_f=3$ flavours, and $\operatorname{Im} 
\Pi_{\rm VV}(q^2)$ is the QCD correlator of vector currents of 
flavour $i$. The electromagnetic spectral function,  
$\operatorname{Im} \Pi^{\rm EM}(s)$, with $s$ the square energy, 
is accessible experimentally from data on $e^+ e^-$ annihilation 
to hadrons as follows. The ratio $R(s)$ is defined as
\begin{equation}
R(s)=
\frac{\sigma_{\rm TOT}(e^{+}e^{-}\rightarrow\text{hadrons})}%
{\sigma(e^{+}e^{-}
\rightarrow\mu^{+}\mu^{-})}\;,
\end{equation}
with
\begin{equation}
\sigma(e^{+}e^{-}\rightarrow \mu^{+}\mu^{-})
=
\frac{4 \pi \alpha_{\rm EM}^{2}}{3s} \;,
\end{equation}
and $\alpha_{\rm EM}=e^{2}/4\pi$. In QCD the ratio $R$ and the 
electromagnetic spectral function are related through
\begin{equation}
R(s) =
12 \,\pi \operatorname{Im} \Pi^{\rm EM}(s) = 
3 \sum_{i=1}^{n_f} Q_i^2 \left( 1 + \frac{\alpha_s}{\pi} 
+ \ldots \right) \, .
\end{equation}
A singlet contribution proportional to $\left(\sum_i Q_i\right)^2$ 
arises at order ${\cal O}(\alpha_s^3)$ and vanishes if one sums 
over three flavours. In particular, for the two-pion final state, 
dominated by the $\rho$-resonance, the relation between $R$ and 
the pion form factor, $F_{\pi}^{(0)}(s)$, is
\begin{equation}
R_{e^{+}e^{-}\rightarrow\pi^{+}\pi^{-}}(s)
= \frac{1}{4} \left(1-\frac{4m_{\pi}^{2}}{s}\right)^{\frac{3}{2}}
\left|F_{\pi}^{(0)}(s)\right|^{2} \,.
\end{equation}
On the QCD side, the vector correlator is assumed to satisfy 
the OPE extended beyond perturbation theory. Non-perturbative 
modifications  due to confinement are parametrized in terms 
of vacuum condensates, i.e.,
\begin{equation}
8\pi \,\Pi_{\rm VV}(Q^2) 
= 
\sum_{N=0}^{\infty} \frac{1}{(Q^2)^N} \,C_{2N}(Q^2,\mu^2) \,
\langle 0| {\cal{O}}_{2N}(\mu^2) |0 \rangle \,,
\end{equation} 
where $Q^2 \equiv - q^2 > 0$ is large, $\mu$ is a renormalization 
scale, and the first term with $N=0$ stands for the PQCD 
contribution. The Wilson coefficients $C_{2N}$, calculable 
in perturbation theory, contain the short distance information 
while the vacuum condensates effectively parametrize the long 
distance dynamics. These condensates are organized according 
to their dimension, with the leading ones being the product of the quark mass and
condensate of dimension $d=4$, $m_q \langle 0| \bar{q} q| 0 \rangle$, 
and the gluon condensate also of $d=4$, $\langle 0|\frac{\alpha_s}{\pi}\,  
G_{\mu\nu}\, G^{\mu\nu}| 0 \rangle$. This gives the result
\begin{equation}
C_4 \langle {\cal{O}}_4 \rangle 
= 
\frac{\pi^2}{3}\, 
\langle \frac{\alpha_s}{\pi}\, G_{\mu\nu}\, G^{\mu\nu} \rangle 
+ 4\, \pi^2 \,\left( m_u \,\langle \bar{u} u \rangle 
+ m_d \,\langle \bar{d} d \rangle + m_s \,\langle \bar{s} s 
\rangle \right)\;,
\end{equation}
where $\alpha_s$ is the running strong coupling, and in the 
sequel $\langle 0| {\cal{O}}_{2N} |0 \rangle \equiv \langle 
{\cal{O}}_{2N}  \rangle $ is to be understood. This condensate 
is renormalization group invariant to all orders in PQCD. 
Next, at dimension $d=6$ there enters the four-quark condensate
\begin{eqnarray}
C_6 \langle {\cal{O}}_6 \rangle 
&=& 
- 4 \pi^3 \langle \alpha_s \left(\bar{u} \gamma_\mu \gamma_5 t^a u 
- \bar{d} \gamma_\mu \gamma_5 t^a d\right)^2 \rangle 
\nonumber 
\\[.2cm]
& & 
- \frac{8}{9} \pi^3 \langle \alpha_s \left(\bar{u} \gamma_\mu t^a u 
+ \bar{d} \gamma_\mu t^a d\right) 
\sum_{q=u,d,s} \bar{q} \gamma_\mu t^a q \rangle 
\;,
\end{eqnarray}
which has a mild dependence on the renormalization scale. 
A once popular approximation is that of vacuum saturation 
\cite{REV} which gives
\begin{equation}
C_6 \langle {\cal{O}}_6 \rangle|_{VS} 
= 
- \frac{896}{81} \, \pi^3 \, \alpha_s 
|\langle \bar{q} \, q \rangle|^2  \simeq - 0.025 \; \mbox{GeV}^6\;. \label{VS}
\end{equation}
This value serves only as an order of magnitude reference, as 
there is no reliable way of estimating corrections to this 
approximation. In our analysis we do not need to invoke vacuum 
saturation as we will be determining the complete $C_6 
\langle {\cal{O}}_6 \rangle$, independently of any theoretical 
assumption. While a dimension $d=2$ contribution is present in the 
PQCD term of the OPE for the vector correlator, it is numerically 
negligible as it is proportional to $m_u^2 + m_d^2$. 

The PQCD vector spectral function is well-known to five-loop 
order 
\cite{Gorishnii:1990vf}-\cite{Celmaster:1979xr}, i.e.
\begin{eqnarray}
8\pi \operatorname{Im}\Pi_{\rm VV}(s) 
& = & 
1 + a_s + a_s^{2}\left(F_{3} + \frac{\beta_1}{2} L_\mu \right)
+ a_s^{3}\left[F_{4} + \left(\beta_1 F_3 
                   + \frac{\beta_2}{2} \right)L_\mu 
+ \frac{\beta_1^2}{4} L_\mu^{2}\right] 
\nonumber \\ [.2cm]
& & 
+ \, a_s^4 \left[k_3 - \frac{\pi^2}{4} \beta_1^2 F_3 
- \frac{5}{24}\pi^2 \beta_1 \beta_2
+ 
\left(\frac{3}{2} \beta_1  F_4 + \beta_2 F_3 
     + \frac{\beta_3}{2}\right) L_\mu 
\nonumber \right. \\ [.2cm]
& & 
+ \left.
\frac{\beta_1}{2} \left(\frac{3}{2} \beta_1 F_3 
                       + \frac{5}{4} \beta_2\right) L_\mu^2
                       + \frac{\beta_1^3}{8} L_\mu^3\right] \;,
\end{eqnarray}
where $a_s \equiv \alpha_s(\mu^2)/\pi$, $L_\mu \equiv 
\ln (Q^2/\mu^2)$, $k_3 = 49.076$ \cite{BCK}, 
$F_3  =  1.9857 - 0.1153 n_f$, 
$F_4  =  18.2427 - \frac{\pi^2}{3} (\frac{\beta_1}{2})^2 
- 4.2158 n_f + 0.0862 n_f^2$, $\beta_1  
=  - \frac{1}{2}(11 - \frac{2}{3}n_f)$,
$\beta_2  =  - \frac{1}{8} (102 - \frac{38}{3}n_f$), 
$\beta_3 = - \frac{1}{32} (\frac{2857}{2} 
- \frac{5033}{18}n_f + \frac{325}{54}n_f^2)$,
and the running coupling to four-loop order is 
\cite{ALPHA}
\begin{eqnarray}
\frac{\alpha^{(4)}_{s}(s_{0})}{\pi} &=&
\frac{\alpha^{(1)}_{s}(s_{0})}{\pi}
+ \Biggl (\frac{\alpha^{(1)}_{s}(s_{0})}{\pi}\Biggr )^{2}
\Biggl (\frac{- \beta_{2}}{\beta_{1}} {\rm ln} L \Biggr ) 
\nonumber 
\\[.4cm]
& & 
+ \Biggl (\frac{\alpha^{(1)}_{s}(s_{0})}{\pi}\Biggr )^{3} 
\Biggl (\frac{\beta_{2}^{2}}{\beta_{1}^{2}} ( {\rm ln}^{2} L -
{\rm ln} L -1) + \frac{ \beta_{3}}{\beta_{1}} \Biggr ) 
\nonumber 
\\[.4cm]
& & 
- \Biggl (\frac{\alpha^{(1)}_{s}(s_{0})}{\pi}\Biggr )^{4}
\Biggl [\frac{\beta_{2}^3}{\beta_{1}^3} \left({\rm ln}^3 L
-\frac{5}{2} {\rm ln}^{2} L - 2 {\rm ln} L + \frac{1}{2}\right) 
\nonumber \\ [.4cm]
& & 
+ \, 3 \frac{\beta_2 \beta_3}{\beta_1^2} {\rm ln} L 
+ \frac{b_3}{\beta_1}
\Biggr ] \; , 
\label{2.9}
\end{eqnarray}
with
\begin{equation}
\frac{\alpha^{(1)}_{s}(s_{0})}{\pi} \equiv\frac{- 2}{\beta_{1} L}\; ,
\label{2.10}
\end{equation}
where $L \equiv\mathrm{ln} (s_{0}/\Lambda_{\overline{\rm MS}}^{2})$ 
defines the standard $\overline{\rm MS}$ scale 
$\Lambda_{\overline{\rm MS}}$, and the constant $b_3$ is
\begin{eqnarray}
b_3&=&\frac{1}{4^4} \Biggl[ \frac{149753}{6} + 3564 \zeta_3
-\left(\frac{1078361}{162} + \frac{6508}{27} \zeta_3 \right) n_F 
\nonumber 
\\
& & 
+ \left(\frac{50065}{162} + \frac{6472}{81} \zeta_3 \right) n_F^2 
+ \frac{1093}{729}
n_F^3 \Biggr ] \; , 
\end{eqnarray}
with $\zeta_{3} = 1.202$. 

Invoking Cauchy's theorem in the complex squared energy $s$-plane 
allows  to relate the experimentally measured hadronic spectral 
function with that from QCD (quark-hadron duality), leading to 
the finite energy sum rules (FESR)
\begin{equation}
(-)^N \, C_{2N+2} \, \langle {\cal{O}}_{2N+2} \rangle 
= 
8 \pi^2 \int_0^{s_0} ds \; s^N \, \frac{1}{\pi}\, 
\mbox{Im} \,\Pi^{\rm DATA}(s) - s_0^{N+1} M_{2N+2}(s_0) \, ,
\label{FESR}
\end{equation}
where the dimensionless PQCD moments $M_{2N+2}(s_0)$ are given by
\begin{eqnarray}
M_{2N+2}(s_0) 
&\equiv& 
- 8 \pi^2  \frac{1}{2 \pi  i} \oint_{C(|s_0|)} \frac{ds}{s_0} 
\left[ \frac{s}{s_0} \right]^N \; \Pi^{\rm PQCD}(s) 
\nonumber 
\\[.2cm]
&=&  
8 \pi^2  \int_0^{s_0} \frac{ds}{s_0} \, 
\left[ \frac{s}{s_0} \right] ^N \, \frac{1}{\pi} \, 
\mbox{Im} \, \Pi(s)^{\rm PQCD}\;.
\end{eqnarray} 
It has been assumed that $s_0$ is large enough so that the 
replacement $\Pi(s) \rightarrow \Pi^{\rm PQCD}(s)$ is justified 
on the circle of radius $|s_0|$ in the complex $s$-plane.
It must be mentioned that there is no mixing of operators of 
different dimension up to second-loop order in the coupling 
\cite{LAUNER}, hence each FESR involves only one condensate. 
We shall neglect radiative corrections to the condensates 
and proceed to determine the condensates of dimension $d=2$ to 
$d=6$ using the FESR, Eq.\ (\ref{FESR}). This approximation is 
justified a-posteriori, given the resulting large errors in the 
condensates.\\
The contour integral in the complex s-plane will be evaluated in the framework of Fixed Order Perturbation Theory (FOPT), as well as in Contour Improved Perturbation Theory (CIPT). In FOPT the coupling and the quark masses at a scale $s_0$ are considered fixed (constant) so that  only logarithmic terms contribute to the integral. The renormalization group (RG) summation of leading logs is only carried out after the contour integration by setting $\mu^2 = - s_0$. In this case the integration can be done analytically. In the case of CIPT the strong coupling and the quark masses are running and the RG is implemented before integrating. The RG equation for the running coupling and quark masses is solved numerically at each point on the circle of radius $s_0$. The value of the strong coupling to be used here is
 $\alpha_s(M_\tau^2) = 0.321 \pm 0.015$, corresponding to $\Lambda_{\overline{\rm MS}} = 341 \pm 24 \; \mbox{MeV}$ in FOPT, or $\alpha_s(M_\tau^2) = 0.344 \pm 0.014$ corresponding to $\Lambda_{\overline{\rm MS}} = 382 \pm 24 \; \mbox{MeV}$ in CIPT \cite{PICH}. If one were to compare 
results for the condensates obtained from $e^+e^-$ data as 
described above with corresponding results from $\tau$-decay, 
one would need to change the overall normalization of the QCD 
correlator.
\begin{table}
\footnotesize
\begin{center}
\begin{tabular}{llll}
\toprule
\noalign{\smallskip}
Number & $e^+e^-\to$ & $\mathcal{I}_0\times 10^{2} (\text{GeV}^2)$ 
                     & $\mathcal{I}_1\times 10^{2} (\text{GeV}^4)$ 
\\
\midrule
1      & $\pi^0\gamma$
                     & $1.51\,(4)(9)$ 
                     & $0.99\,(3)(6)$ 
\\
2      & $\eta\gamma$ 
                     & $0.30\,(4)(2)$ 
                     & $0.29\,(5)(2)$ 
\\
3      & $\pi^+\pi^-$
                     & $133.1\,(25)(8)$ 
                     & $95.6\,(27)(7)$ 
\\
4      & $\pi^+\pi^-\pi^0$ 
                     & $30.1\,(3)(14)$ 
                     & $41.6\,(8)(16)$ 
\\
5      & $2(\pi^+\pi^-)$ 
                     & $60.5\,(3)(30)$ 
                     & $156.6\,(10)(78)$ 
\\
6      & $\pi^+\pi^-2(\pi^0)$ 
                     & $79.6\,(4)(69)$ 
                     & $206.6\,(13)(181)$ 
\\ 
7      & $2\pi^+2\pi^-\pi^0$ 
                     & $7.1\,(2)(7)$ 
                     & $20.8\,(6)(20)$ 
\\ 
8      & $3(\pi^+\pi^-)$ 
                     & $2.59\,(7)(2)$ 
                     & $8.83\,(30)(60)$ 
\\
9      & $2(\pi^+\pi^-\pi^0)$ 
                     & $11.6\,(3)(12)$ 
                     & $39.2\,(10)(41)$ 
\\
10     & $\eta(\pi^+\pi^-)$ 
                     & $6.9\,(3)(6)$ 
                     & $18.9\,(8)(16)$ 
\\ 
11     & $\eta\omega$ 
                     & $3.1\,(1)(3)$ 
                     & $9.0\,(3)(9)$ 
\\ 
12     & $\eta (2\pi^+2\pi^-)$ 
                     & $1.0\,(1)(1)$ 
                     & $3.6\,(4)(3)$ 
\\ 
13     & $\omega\pi^0(\omega\to \pi^0\gamma)$ 
                     & $2.22\,(4)(6)$ 
                     & $4.5\,(1)(1)$ 
\\
14     & $K^+K^-$    & $25.2\,(3)(6)$ 
                     & $34.0\,(7)(9)$ 
\\
15     & $K_{s}^0K_{L}^0$ 
                     & $12.2\,(2)(5)$ 
                     & $14.2\,(5)(7)$ 
\\  
16     & $\eta \phi$ & $3.6\,(2)(3)$ 
                     & $11.5\,(7)(8)$ 
\\  
17     & $p\bar{p}$  & $0.82\,(3)(4)$ 
                     & $3.0\,(1)(2)$ 
\\
18     & $n\bar{n}$  & $0.9\,(3)(1)$ 
                     & $3.5\,(13)(5)$ 
\\ 
19     & Inclusive   & $56.0\,(4)(2)$ 
                     & $231.6\,(30)(9)$ 
\\ 
\\ 
20*    & $\pi^+\pi^-$ (CHPT) 
                     & $0.01(0)(0)$ 
                     & $0.001(0)(0)$ 
\\
21*    & $\pi^+\pi^-3\pi^0$ 
                     & $3.5\,(1)(12)$ 
                     & $10.4\,(3)(34)$ 
\\
22*    & $\pi^+\pi^-(4\pi^0)$ 
                     & $1.9\,(1)(6)$ 
                     & $6.2\,(2)(21)$ 
\\
23*    & $\omega \pi^+\pi^-$, $\omega2\pi^0(\omega\to \pi^0\gamma)$ 
                     & $0.56\,(2)(18)$ 
                     & $1.6\,(1)(5)$ 
\\ 
24*    & $\omega (\text{non-}3\pi,\pi \gamma,\eta\gamma)$ 
                     & $0.11\,(1)(4)$ 
                     & $0.07\,(1)(2)$ 
\\ 
25*    & $\phi (\text{non-}K\overline{K}, 3\pi,\pi \gamma,\eta\gamma)$ 
                     & $0.04\,(0)(1)$ 
                     & $0.04\,(0)(1)$ 
\\ 
26*    & $\eta \pi^+\pi^-(2\pi^0)$ 
                     & $1.0\,(1)(3)$ 
                     & $3.6\,(4)(12)$ 
\\ 
27*    & $K\overline{K}\pi$ 
                     & $16.6\,(4)(57)$ 
                     & $47\,(12)(16)$ 
\\ 
28*    & $K\overline{K}2\pi$ 
                     & $36.1\,(9)(64)$ 
                     & $124\,(3)(22)$ 
\\ 
29*    & $K\overline{K}3\pi$ 
                     & $1.7\,(1)(3)$ 
                     & $6.1\,(5)(10)$ 
\\ 
\\
       & Total       & $499\,(3)(13)$ 
                     & $1105\,(5)(38)$ 
\\
\bottomrule
\end{tabular} 
\end{center}
\caption{\small
The contributions to $\mathcal{I}_n=\frac{1}{2}\int^{s_0}_{0}s^n 
R(s)\,ds$, using $s_0=4.5\,\text{GeV}^2$. The first error is 
statistical, while the second is systematic. For the total 
uncertainty, we added the uncertainties of all the final 
states in quadrature. Stars ($*$) indicate that the cross 
section had to be estimated, following the prescriptions 
explained in Appendix \ref{app-data}.
}
\label{tab-contrib}
\end{table}
\begin{figure}[t!]
\begin{center}
\includegraphics[scale=0.80]{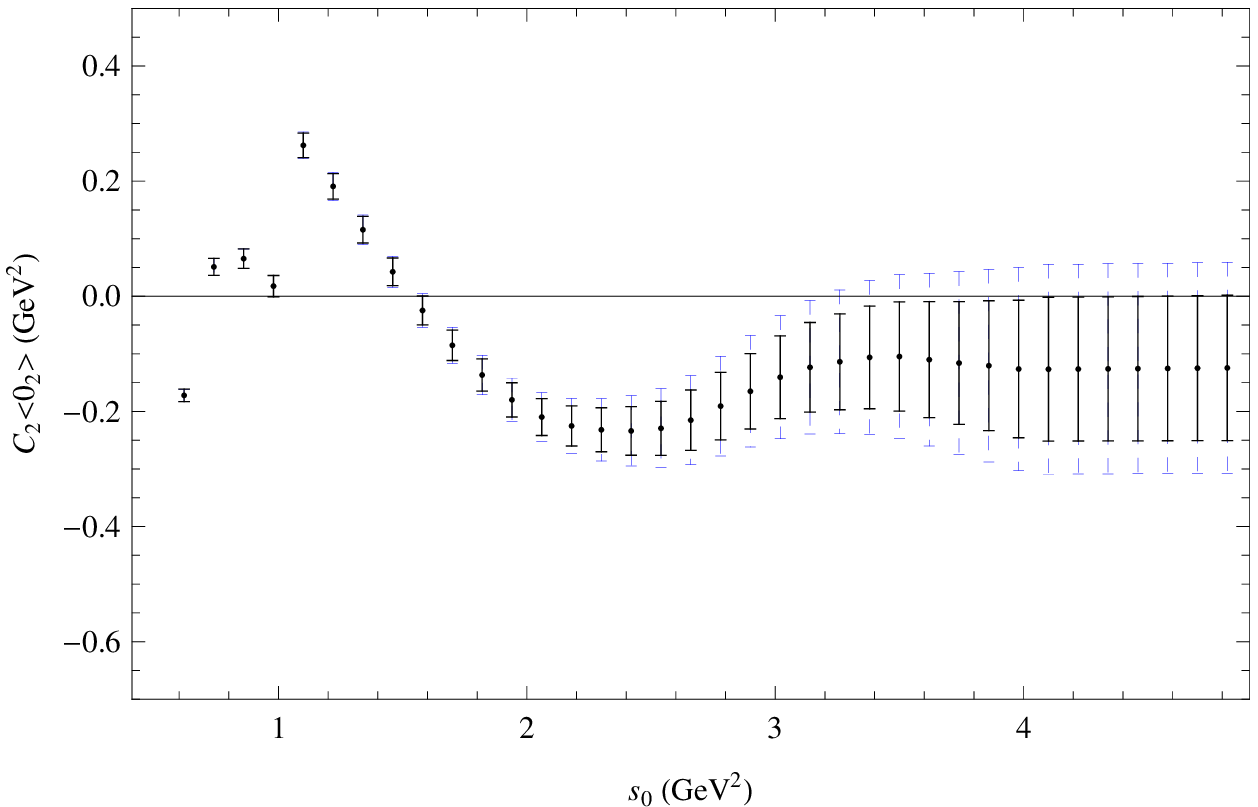}
\end{center}
\caption{\footnotesize{
$C_2\left<\mathcal{O}_2\right>$ calculated in FOPT using $\alpha_s(M_\tau^2) = 0.321 \pm 0.015$, corresponding to 
$\Lambda_{\overline{\rm MS}}^{(n_f=3)} = 341$ MeV. The smaller  
uncertainties are obtained assuming no correlations between 
experiments, while the larger ones assume $100\,\%$ correlations 
for data obtained using the same experimental facility.
}}
\label{fig1}
\end{figure}
\begin{figure}[ht]
\begin{center}
\includegraphics[scale=0.80]{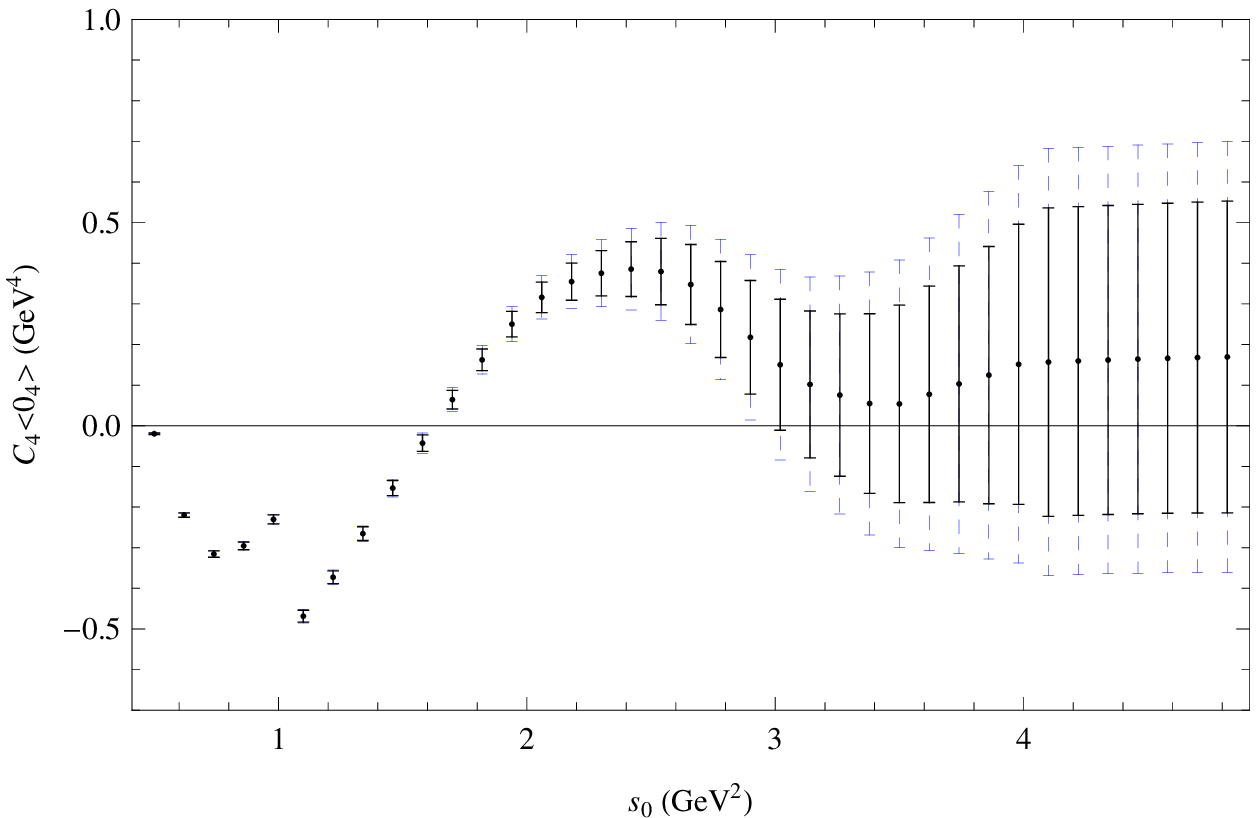}
\end{center}
\caption{\footnotesize{
$C_4\left<\mathcal{O}_4\right>$ calculated in FOPT using $\alpha_s(M_\tau^2) = 0.321 \pm 0.015$, corresponding to 
$\Lambda_{\overline{\rm MS}}^{(n_f=3)} = 341$ MeV. The smaller 
uncertainties are obtained assuming no correlations between 
experiments, while the larger ones assume $100\,\%$ correlations 
for data obtained using the same experimental facility.
}}
\label{fig2}
\end{figure}
\section{Hadronic Data}
The most rigorous approach to making a collection of exclusive 
data is to combine all available data for a given final state 
\cite{etaak}-\cite{KSX}. Their respective statistical and 
systematic uncertainties should be used to define the weight 
in the averaging procedure. In fact, this is important for 
precision determinations, such as, e.g., the  $g-2$ factor of 
the muon. However, one can be less rigorous for the purpose of 
the present analysis. The philosophy is to make use of only the 
most recent data for a given exclusive hadronic final state. If 
the most recent data do not cover the entire energy range for 
that final state, then older data are used to cover this region. 
In this way there are no overlapping data sets, and then one 
does not need to be concerned about how precisely different, 
often inconsistent, data are combined. We will make use of 
exclusive data up to $\sqrt{s} = 2$ GeV, and above this energy 
we use the BES inclusive data \cite{bes2002}-\cite{bes2009}. These 
data are consistent with PQCD, hence one observes a plateau 
for the condensates above $\sqrt{s_0} = 2$ GeV. Since, in 
addition, their systematic errors are small (3\,\%) we observe 
that the resulting errors for $C_N\left<\mathcal{O}_N\right>$ 
do not change any more above this energy.
The treatment of the statistical uncertainties is standard. We 
assume no correlations either within individual experiments or 
between different experiments. The exception to this is the 
treatment of the $\pi^+\pi^-$ data from BaBar \cite{2pibabar}. 
In this case we are given a full correlation matrix. To simplify 
the calculation, we simply assumed $100\,\%$ correlation for the 
statistical uncertainty, which gives an upper bound on the  
uncertainty.\footnote{We also checked that this assumption does
  not have a significant impact on the uncertainty in the final 
  results for the condensates.} 
For the systematic errors, we assume that they are completely 
correlated within a data set. It is reasonable to assume that 
there will also be correlations between different final-state 
data sets obtained by the same experimental facility, e.g.\ 
BaBar, CMD2 or SND, since each experiment would use the same 
procedure when applying radiative corrections or for the 
determination of the luminosity. As we do not have any information 
about these correlations, we will quote our final uncertainty 
both for the worst-case scenario assuming a $100\,\%$ correlation 
of data determined at the same facility, as well as for the 
best-case scenario of no correlations. The true uncertainty 
can be taken to lie between these two cases. For more details 
see the Appendix where we also present the list of all final 
states used in our analysis in Table \ref{tab-datasets}.
We have verified our data base against the one used in 
\cite{davier2011} for the determination of the muon anomaly, 
$a_{\mu}$. 
 
\section{Results and Conclusions}
\begin{figure}[t!]
\begin{center}
\includegraphics[scale=0.80]{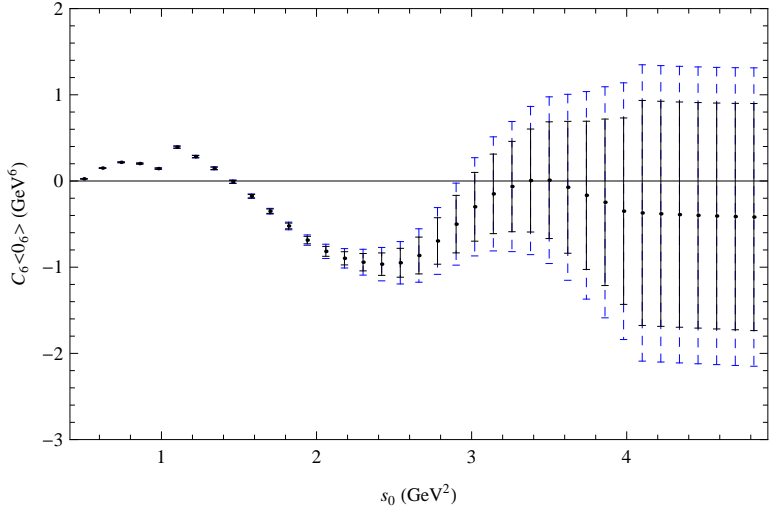}
\end{center}
\caption{\footnotesize{
$C_6\left<\mathcal{O}_6\right>$ calculated in FOPT using $\alpha_s(M_\tau^2) = 0.321 \pm 0.015$, corresponding to 
$\Lambda_{\overline{\rm MS}}^{(n_f=3)}=341$ MeV. The smaller 
uncertainties are obtained assuming no correlations between 
experiments, while the larger ones assume $100\,\%$ correlations 
for data obtained using the same experimental facility.
}}
\label{fig3}
\end{figure}
Results for the condensates of the total $I=0$ plus $I=1$ current 
correlator determined from $e^+e^-$ data are presented in Figs.\ 
\ref{fig1}-\ref{fig3}. From standard QCD and the OPE one would 
expect the dimension $d=2$ condensate to be zero. Fig.\ \ref{fig1} 
shows that this is the case within (large) errors, although the 
central value is negative. It is also seen from this figure 
that the sum rule starts to be saturated only at about $s_0 
\approx 3.4$ GeV$^{2}$. For the condensate of dimension $d=4$ 
shown in Fig.\ \ref{fig2} we observe saturation of the sum rule 
at a slightly higher $s_0$ than in the case of dimension $d=2$. 
The central value of the dimension $d=4$ condensate is seen to 
be mostly positive, as it should be since it is related to the 
vacuum energy. Figure \ref{fig3} shows the result for the 
dimension $d=6$ condensate. Since the errors are so large, no 
meaningful conclusion can be drawn in this case, other than that 
the central value is negative, in agreement with the vacuum 
saturation approximation, Eq.\ (\ref{VS}).
The results for the condensates are not very sensitive to the 
value of $\alpha_{s}$ used in the sum rules. Their values at  
$s_{0} = 4.5$ GeV$^{2}$, assuming uncorrelated data errors
are
\begin{eqnarray}
C_{2}\left\langle \mathcal{O}_{2}\right\rangle = \;\Bigg\{ 
\begin{array}{lcl}
(-0.13 \pm 0.13|_{\text{EXP}} \pm 0.03|_{\alpha_{s}}) \,
  \text{GeV}^{2} |_{\text{FOPT}}\; \\ [.3cm]
(-0.14 \pm 0.13|_{\text{EXP}} \pm 0.03|_{\alpha_{s}}) \,
  \text{GeV}^{2} |_{\text{CIPT}}\; \label{C2a} \;
\end{array}
\end{eqnarray}
\begin{eqnarray}
C_{4}\left\langle \mathcal{O}_{4}\right\rangle = \;\Bigg\{
\begin{array}{lcl}
 (+0.16 \pm 0.38|_{\text{EXP}} \pm 0.04|_{\alpha_{s}}) \,
  \text{GeV}^{4}|_{\text{FOPT}} \;  \\[.3cm]
  (+0.10 \pm 0.38|_{\text{EXP}} \pm 0.04|_{\alpha_{s}}) \,
  \text{GeV}^{4}|_{\text{CIPT}} \;\label{C4a}
\end{array} 
\end{eqnarray}
\begin{eqnarray}
C_{6}\left\langle \mathcal{O}_{6}\right\rangle = \;\Bigg\{
\begin{array}{lcl}
 (-0.40 \pm 1.30|_{\text{EXP}} \pm 0.10|_{\alpha_{s}}) \,
  \text{GeV}^{6}|_{\text{FOPT}} \;  \\[.3cm]
  (-0.40 \pm 1.30|_{\text{EXP}} \pm 0.10|_{\alpha_{s}}) \,
  \text{GeV}^{6}|_{\text{CIPT}} \;\label{C6a}
\end{array} 
\end{eqnarray}
Assuming instead maximally correlated data errors the experimental uncertainties above increase to $\pm 0.18|_{\text{EXP}}$, $\pm 0.52|_{\text{EXP}}$, and $\pm 1.70|_{\text{EXP}}$, for $C_{2}\left\langle \mathcal{O}_{2}\right\rangle$, $C_{4}\left\langle \mathcal{O}_{4}\right\rangle$, and $C_{6}\left\langle \mathcal{O}_{6}\right\rangle$, respectively. 
\begin{table}
\footnotesize
\begin{center}
\begin{tabular}{cllcl}
\toprule
\noalign{\smallskip}
Number & $e^+e^-\to$ & Reference & Data Range & Vacuum Pol.
\\
       &             &           & $\sqrt{s}$ (GeV) & 
\\
\midrule
1      & $\pi^0\gamma$ & CMD-2 (2005) \cite{etaak}
                                 & $0.6-1.31$ & Dressed 
\\
2      & $\eta\gamma$ & CMD-2 (2005) \cite{etaak}  
                                 & $0.6-1.38$ 	& Dressed 
\\
3      & $\pi^+\pi^-$ & BaBar (2009) \cite{2pibabar} 
                                 & $0.31-2.95$ & Bare 
\\
4      & $\pi^+\pi^-\pi^0$ & BaBar (2004) \cite{2004babar} 
                                 & $1.06-2.99$ & Dressed 
\\
       &               & CMD-2 (2006) \cite{cmd22006} 
                                 & $1.01-1.06$ & Dressed 
\\
       &               & SND (2002) \cite{ach2002} 
                                 & $0.98-1.01$ & Dressed 
\\
       &               & SND (2003) \cite{ach2003} 
                                 & $0.66-0.98$ & Dressed 
\\
5      & $2(\pi^+\pi^-)$ & BaBar (2005) \cite{4pi1babar} 
                                 & $0.62-4.45$ & Bare 
\\
6      & $\pi^+\pi^-2(\pi^0)$ & BaBar (2010) \cite{babarpre} 
                                 & $0.76-3.31$ & Dressed 
\\ 
       &               & SND (2009) \cite{4piachasov} 
                                 & $0.66-0.94$ & Dressed 
\\
7      & $2\pi^+2\pi^-\pi^0$ & BaBar (2007) \cite{etababar} 
                                 & $1.03-2.98$ & Dressed 
\\ 
8      & $3(\pi^+\pi^-)$ & BaBar (2006) \cite{6pibabar} 
                                 & $1.3-4.5$ & Dressed
\\
9      & $2(\pi^+\pi^-\pi^0)$ & BaBar (2006) \cite{6pibabar} 
                                 & $1.3-4.5$ & Dressed 
\\
10     & $\eta(\pi^+\pi^-)$ & BaBar (2007) \cite{etababar} 
                                 & $1.03-2.98$ & Dressed 
\\ 
11     & $\eta\omega$ & BaBar (2006) \cite{6pibabar} 
                                 & $1.25-3.25$ & Dressed 
\\ 
12     & $\eta (2\pi^+2\pi^-)$ & BaBar (2007) \cite{etababar} 
                                 & $1.31-2.89$	 & Dressed 
\\ 
13     & $ \omega\pi^0(\omega\to \pi^0\gamma)$ & CMD-2 (2003) \cite{2pi0ak} 
                                 & $0.92-1.38$ & Bare 
\\
14     & $ K^+K^-$ & CMD-2 (2008) \cite{akKK} 
                                 & $1.011-1.034$ & Dressed 
\\
       &           & SND (2007) \cite{ach2007} 
                                 & $1.04-1.38$	 & Dressed 
\\
       &           & DM2 (1988) \cite{KKbisello} 
                                 & $1.38-2.40$	 & ? 
\\
15     & $ K_{s}^0K_{L}^0$ & SND (2001) \cite{ach2001} 
                                 & $1.01-1.06$ & Dressed 
\\  
       &           & SND (2006) \cite{ach2006} 
                                 & $1.04-1.38$ & Dressed 
\\
       &           & DM1 (1981) \cite{KLKsdm1} 
                                 & $1.4-2.1$ & ? 
\\
16     & $\eta \phi$ & BaBar (2008) \cite{Kbabar} 
                                 & $1.57-3.45$ & Dressed 
\\  
17     & $p\bar{p}$ & BaBar (2006) \cite{babarpp} 
                                 & $1.88-4.2$ & Dressed 
\\
18     & $n\bar{n}$ & Fenice (1998) \cite{nn} 
                                 & $1.9-2.44$ & Dressed 
\\
19     & Inclusive & BES (2002) \cite{bes2002} 
                                 & $2-5$ & Bare 
\\
       &           & BES (2009) \cite{bes2009} 
                                 & $2.6-3.65$ & Bare 
\\ 
\\
20     & $\pi^+\pi^-$ ($\chi$pT) & *
                                 & $0.14-0.31$ & 
\\
21     & $\pi^+\pi^-3\pi^0$      & * 
                                 & $1.03-2.98$	 & 
\\
22     & $\pi^+\pi^-(4\pi^0)$    & * 
                                 & $1.3-4.5$ &	 
\\
23     & $\omega \pi^+\pi^-$, $\omega2\pi^0(\omega\to \pi^0\gamma)$ 
                                 & * 
                                 & $1.15-2.53$	 & 
\\ 
24     & $\omega (\text{non-}3\pi, \pi \gamma,\eta\gamma)$ 
                                 & * 
                                 & $0.7-0.8$ & 
\\ 
25     & $\phi (\text{non-}K\overline{K}, 3\pi,\pi \gamma,\eta\gamma)$ 
                                 & * 
                                 & $1.01-1.03$ & 
\\ 
26     & $\eta \pi^+\pi^-(2\pi^0)$ 
                                 & * 
                                 & $1.3-2.9$ & 
\\ 
27     & $K\overline{K}\pi$
                                 & * 
                                 & $1.3-4.7$ & 
\\ 
28     & $K\overline{K}2\pi$
                                 & * 
                                 & $1.4-4.3$ & 
\\ 
29     & $K\overline{K}3\pi$ 
                                 & * 
                                 & $1.6-4.5$ & 
\\ 
\bottomrule
\end{tabular} 
\end{center}
\caption{\small
Data used in evaluating the condensates. Stars ($*$) indicate 
that the final state has not been measured and was estimated 
as explained in the text. The treatment of the vacuum 
polarization correction is described in the text. Question 
marks indicate that it is not known which corrections have 
been applied by the experimental collaboration, and no 
corrections were applied by us in these cases.
}
\label{tab-datasets}
\end{table}
These results are consistent with expectations and with the 
values extracted from $\tau$-decay 
\cite{C2_1}-\cite{C2_4}, albeit within 
very large errors. The uncertainties in all the results are clearly dominated by the experimental errors.
One consequence of this is that e.g. in the $d=2$ case it is not possible to distinguish among various theoretical sources of such a term in the OPE. This is also the case with the result  from the $\tau$ decay data \cite{C2_1}-\cite{Dominguez:2006ct}, although  it has a smaller uncertainty. In any case, the impact of the PQCD input for a given integration method (FOPT or CIPT) is such that the higher the value of $\alpha_s$ the lower the curve in Fig.\ref{fig1}. A similar trend is found from additional higher order PQCD terms, and this behaviour is common to all condensates. For the gluon condensate the value determined 
from charmonium data is $C_4\langle {\cal{O}}_4 \rangle = 0.06 
\pm 0.04 \, \mbox{GeV}^4$ \cite{REV}, \cite{REV2}-\cite{Bein}, where the error is our 
conservative estimate. The result from $\tau$-decay is \cite{Dominguez:2006ct} $C_4\langle {\cal{O}}_4 \rangle = 0.07 \pm 0.02 \mbox{GeV}^4$ for a similar input value of $\alpha_s$.
Concerning the $d=6$ result, given the large uncertainty in its value no meaningful comparison can be made with e.g. the vacuum saturation prediction, Eq.(\ref{VS}).
It should be emphasized that the consistency between data and 
QCD is in no way obvious. The high sensitivity of the FESR can 
be demonstrated if one were to determine condensates from the 
individual $I=0$ and $I=1$ channels separately. For example, 
for the non-strange $I=1$ final states listed in lines 2, 3, 5, 
6, 8-10, 12, 22, and 26 of Tables \ref{tab-contrib} and 
\ref{tab-datasets}, the dimension $d=2$ and $d=4$ condensates would
turn out  to be unacceptably large, i.e.  more than $3\sigma$ away from 
theoretical expectations. Moreover, with this subset of data 
there would no longer be a clear plateau in the $s_0$-dependence , as the one 
 in Figs.\ \ref{fig1}, \ref{fig2} and \ref{fig3}. 
We believe that these unacceptable results indicate that the 
separation of the data into $I=1$ and $I=0$ components 
requires a fairly detailed understanding of isospin symmetry 
breaking. If isospin symmetry is broken, states with different 
isospin quantum numbers can mix and unknown interference 
contributions can spoil a naive decomposition into a sum of 
isovector plus isoscalar parts.\\

To summarize, with the help of FESR we have confronted QCD and 
the OPE with experimental data for $R(s)$. In particular we have 
considered the zeroth, first and second moment of $R(s)$. This 
comparison of theory with experiment can simultaneously help to 
give support to the theoretical framework, i.e.\ the basic 
principles of QCD, the OPE and analyticity, and as a test of the 
quality and completeness of the data. 
We found that recent $e^{+}e^{-}$ annihilation data 
are consistent with QCD and the OPE within very large errors, 
even though there exists some indication that not all data are 
yet correctly accounted for. Any future improvement on the accuracy 
of vacuum condensate determinations from $e^+ e^-$ annihilation 
will require a considerable reduction in the experimental 
uncertainties.  
\appendix
\section{Hadronic Data}
\label{app-data}

In this Appendix we give details of the data sets used in our 
analysis. Table~\ref{tab-datasets} lists all relevant final 
states, their corresponding energy ranges and the corresponding  
references.
Some of the exclusive channels are as yet unmeasured. We give here 
the estimated values of these missing channels. These estimates 
are mostly based on isospin arguments, primarily as described in 
Refs.\ \cite{davier2011} and \cite{davier2003}, where further 
details can be found. It should be mentioned that the Pais 
isospin-class analysis \cite{davier2011} can be employed only 
after removing the $\eta$-contribution from some of the final 
states. This also avoids double counting in the evaluation of 
the FESR integrals. We have only quoted the results. A $33\,\%$ 
model error has been assigned to all of these data, unless stated 
otherwise. The numbers in the following list refer to the final 
states listed in Table \ref{tab-datasets}.

\begin{enumerate}
\item[20.]
There are no data for the $\pi^+\pi^-$ final state at low energies, 
so the prediction of chiral perturbation theory (CHPT) has been 
used. The cross section is expressed in terms of the $\pi$ form 
factor, $\sigma( \pi^+\pi^-) = \pi\alpha^2\beta_{0}^{3} 
|F_{\pi}^{0}|^2/(3s)$, where $\beta_0 = (1-4 m_{\pi}^{2}/s)^{1/2}$. 
The form factor is taken of the form $F_{\pi}^{0} = 1 + 
\frac{1}{6} \langle r^2\rangle_\pi s+c_1 s^2+c_2 s^3 + 
\mathcal{O}(s^4)$. The constants are found by fitting 
the form factor to space-like data, with the result 
$\langle r^2\rangle_\pi = (0.439\pm 0.008) \, \text{fm}^2$, 
$c_1 = (6.8\pm 1.9) \, \text{GeV}^{-4}$ and 
$c_2 = (-0.7\pm 6.8) \, \text{GeV}^{-8}$ \cite{davier2003}. 
\item[21.]
A prediction for the missing channel 
$\sigma(\pi^+\pi^-3\pi^0)_{\eta-\text{excl}.} 
= \frac{1}{2}\sigma(2\pi^+2\pi^-\pi^0)_{\eta-\text{excl}.}$ 
can be obtained after removing the $\eta$-contribution: 
$\sigma(2\pi^+2\pi^-\pi^0)_{\eta-\text{excl}.} = 
\sigma(2\pi^+2\pi^-\pi^0)-\sigma(\eta\pi^+\pi^-) \times 
\mathcal{B}(\eta\to\pi^+\pi^-\pi^0)$, where 
$\mathcal{B}(\eta\to\pi^+\pi^-\pi^0)=0.2274\pm 0.0028$ and 
where $\sigma(2\pi^+2\pi^-\pi^0)$ is obtained from BaBar (2007) 
\cite{etababar}. 
\item[22.]
$\sigma(\pi^+\pi^-4\pi^0) = 0.0625\, \sigma(3\pi^+3\pi^-) +
0.145\, \sigma(2\pi^+2\pi^-2\pi^0)_{\eta-\text{excl}}$ from 
isospin symmetry, where the $\eta$-subtracted cross section 
$\sigma(2\pi^+2\pi^-2\pi^0)_{\eta-\text{excl}}$ is determined 
using $\sigma(2\pi^+2\pi^-2\pi^0)_{\eta-\text{excl}.} = 
\sigma(2\pi^+2\pi^-2\pi^0)-\sigma(\eta\omega) \times 
\mathcal{B}(\eta\to\pi^+\pi^-\pi^0)\times 
\mathcal{B}(\omega\to\pi^+\pi^-\pi^0)$, where 
$\mathcal{B}(\omega\to\pi^+\pi^-\pi^0)=0.892\pm 0.007$, 
$\sigma(2\pi^+2\pi^-2\pi^0)$ was measured by BaBar 
\cite{6pibabar}, and $\sigma(\eta\omega)$ was taken from 
\cite{6pibabar}.
\item[23.]
One has  $\sigma(\omega \pi^+\pi^-$) from BaBar (2007) 
\cite{etababar}. However, the dominant three-pion decay of the 
$\omega$ already appears in the five-pion final state. Thus we 
calculate the contribution for $\omega\to \pi^0\gamma$ by using 
$\sigma(\omega \pi^+\pi^-\to \pi^+\pi^-\pi^0 \gamma) 
= \sigma(\omega \pi^+\pi^-)\times \mathcal{B}(\omega\to \pi^0\gamma)$ where $\mathcal{B}(\omega\to \pi^0\gamma)=0.0828\pm 0.0028$. 
Then from isospin arguments it follows that $\sigma(\omega 2\pi^0\to 3\pi^0 
\gamma)=0.5\,\sigma(\omega \pi^+\pi^-\to \pi^+\pi^-\pi^0 \gamma)$.
\item[24.]
There are cross section data for $\sigma(\omega\to \pi^+\pi^-\pi^0)$ 
from \cite{cmd22004}. This is the major decay mode of the 
$\omega$, with $\mathcal{B}(\omega\to \pi^+\pi^-\pi^0)=0.892\pm 
0.007$. In addition to this final state, we have already accounted 
for the $\pi^0\gamma$, $\pi^+\pi^-$, and $\eta \gamma$ final states. 
We thus estimate the contribution from $\omega$ decays not yet 
accounted for as being 
$\sigma(\omega\to \pi^+\pi^-\pi^0)\times 
\mathcal{B}(\text{non}-3\pi,\pi \gamma,\eta\gamma)/ 
\mathcal{B}(\omega\to \pi^+\pi^-\pi^0)$, where 
$\mathcal{B}(\text{non}-3\pi,\pi \gamma,\eta\gamma)=0.0094$.  
\item[25.]
We have already accounted for $\phi$ to $K\overline{K}$, $3\pi$, 
$\pi\gamma $ and $\eta \gamma$. Hence there is a missing contribution 
 found from $\mathcal{B}(\text{missing}) = 
1-\mathcal{B}(\phi\to K\overline{K})-\mathcal{B}(\phi\to 3\pi) - 
\mathcal{B}(\phi\to \pi \gamma)-\mathcal{B}(\phi\to\eta \gamma)
=0.0014$. We then calculate $\sigma(\phi (\text{non-}K\overline{K},
3\pi,\pi \gamma,\eta\gamma))=\sigma(\phi\to K^+K^-) 
\mathcal{B}(\text{missing})/\mathcal{B}(\phi\to K^+K^-) = 
0.003\,\sigma(\phi\to K^+K^-)$.
\item[26.]
We assume that  $\sigma(\eta 2\pi^+2\pi^-)=\sigma(\eta \pi^+
\pi^-2\pi^0)$.
\item[27.]
One has $\sigma(K\overline{K}\pi)=3\sigma(K_{s}^{0}K^\pm\pi^\mp) +
\sigma(\phi\pi^0)\times \mathcal{B}(\phi\to K\overline{K})$. 
Here $\sigma(K_{s}^{0}K^\pm\pi^\mp)$ and $\sigma(\phi\pi^0)$ 
are obtained from BaBar (2008) \cite{Kbabar}. In addition,
$\mathcal{B}(\phi\to K\overline{K})=0.831 \pm 0.003$.  
\item[28.]
$\sigma(K\overline{K}2\pi)=9\sigma(K^+K^-2\pi^0) + 
\frac{9}{4}\sigma(K^+K^-\pi^+\pi^-)$. The cross sections for 
the $K^+ K^- 2\pi$ final states are available from BaBar 
(2007) \cite{Babar2007b}. To estimate an uncertainty for this 
result we make use of a different procedure from 
\cite{Hagiwara:2003da} which uses the inclusive data $K_SX$ 
(DM1 Collaboration, \cite{KSX}) as a starting point. The 
difference in cross sections between these methods is $17\%$, 
which we take to be the systematic uncertainty.
\item[29.]
We assume that $\sigma(K^0\overline{K}^0\pi^+\pi^-\pi^0)_{\eta - 
\text{excl}}=\sigma(K^+K^-\pi^+\pi^-\pi^0)_{\eta-\text{excl}}$. 
With these two processes as primary contributors we find 
$\sigma(K\overline{K}3\pi)=2\sigma(K^+K^-\pi^+\pi^-\pi^0)_{\eta 
- \text{excl}}$.  
We calculate $\sigma(K^+K^-\pi^+\pi^-\pi^0)_{\eta-\text{excl}} = 
\sigma(K^+K^-\pi^+\pi^-\pi^0)-\sigma(\phi\eta)\times 
\mathcal{B}(\phi\to K^+K^-)\times \mathcal{B}(\eta\to 
\pi^+\pi^-\pi^0)$, where  $\mathcal{B}(\phi\to K^+K^-) = 
0.489\pm 0.01$ and $\mathcal{B}(\eta\to \pi^+\pi^-\pi^0) = 
0.2274\pm 0.0028$. We obtain $\sigma(K^+K^-\pi^+\pi^-\pi^0)$ 
from BaBar (2007) \cite{etababar}, and $\sigma(\phi\eta)$ is 
taken from BaBar (2008) \cite{Kbabar}. 
\end{enumerate}

When using the cross sections in the sum rules one needs the 
so-called {\it undressed} (or {\it bare}) cross-sections, 
$\sigma_{0}$. The bare cross sections are obtained by removing 
initial-state radiation, but not final-state radiation, and the 
contribution due to vacuum polarization. The leptonic part of 
the vacuum polarization has been subtracted already from the 
BaBar, SND, and CMD2 data. However, most of the data have not 
been corrected for the hadronic vacuum polarization. We do this 
by using 
\begin{equation}
\sigma_{0} = 
\left(\frac{\alpha(0)}{\alpha(s)}\right)^2\sigma_{\text{born}} = 
|1-\Pi'(s)|^2\sigma_{\text{born}}
\end{equation}
where $\Pi'(s)$ is the vacuum polarization function 
$\Pi'(s)\equiv \Pi(s)-\Pi(0)$ and $\Pi(s)$ is obtained from a 
dispersion relation integrating over hadronic data. 

\vspace*{8mm}

{\bf \Large Acknowledgments}

We thank Miriam Fritsch and Achim Denig for discussions about 
experimental aspects of $e^+e^-$ data, in particular those of 
Ref.\ \cite{babarpre}. SE acknowledges the kind hospitality at the University of Mainz, and partial support by RFBR grants 10-02-695, 11-02-112, and 11-02-558.


\end{document}